# The relative influences of government funding and international collaboration on citation impact

Loet Leydesdorff,*[1] Lutz Bornmann,[2] and Caroline S. Wagner [3]

**Abstract**

In a recent publication in *Nature,* Wagner & Jonkers (2017) report that public R&D funding is only weakly correlated with the citation impact of a nation's papers as measured by the field-weighted citation index (FWCI; defined by Scopus). On the basis of the supplementary data, we upscaled the design using Web-of-Science data for the decade 2003-2013 and OECD funding data for the corresponding decade assuming a two-year delay (2001-2011). Using negative binomial regression analysis, we find very small coefficients, but the effects of international collaboration are positive and statistically significant, whereas the effects of government funding are negative, an order of magnitude smaller, and statistically non-significant (in two of three analyses). In other words, international collaboration improves the impact of average research papers, whereas more government funding tends to have a small adverse effect when comparing OECD countries.

[1] *corresponding author; Amsterdam School of Communication Research (ASCoR), University of Amsterdam PO Box 15793, 1001 NG Amsterdam, The Netherlands; loet@leydesdorff.net
[2] Division for Science and Innovation Studies, Administrative Headquarters of the Max Planck Society, Hofgartenstr. 8, 80539 Munich, Germany; bornmann@gv.mpg.de
[3] John Glenn College of Public Affairs, The Ohio State University, Columbus, Ohio, USA, 43210; wagner.911@osu.edu

**Introduction**

To view the national impact of international collaboration, Wagner & Jonkers (2017) assigned papers and impact measures to countries using fractional counting and a field-weighted citation index (FWCI), as defined by the Scopus team at Elsevier (Plume & Kamalski, 2014). They found "a clear correlation between a nation's scientific influence and the links it fosters with foreign researchers." The authors show that public R&D funding is only weakly correlated with the citation impact of a nation's papers. To reach this conclusion, the authors created an index of openness with values assigned for OECD countries. The data is available for download at go.nature.com/2fzrnt3.

The Comment in *Nature* remains at the level of pair-wise correlations. In our opinion, this data allows for a next step: the relative influences of government funding and international collaborations on citation impact can be tested using regression analysis. Has international collaboration in the meantime become an independent factor in the self-organization of the sciences (Persson, Glänzel, & Danell, 2004; Wagner & Leydesdorff, 2005)? Or is domestic stimulation by national governments a more crucial factor? It has been argued that the sciences are self-organizing, and thus relatively resilient against changes in external funding priorities by governments (van den Daele & Weingart, 1975; cf. van den Besselaar & Sandström, 2017).

To test the hypothesis further, we scaled up to a decade of data (2003-2013) using the funding data (Government Budget Allocations for R&D; GBARD)[1] of 35 OECD member states and seven affiliated economies,[2] on the one side, and using our access to an in-house version of the Web-of-Science (WoS) developed and maintained by the Max Planck Digital Library (MPDL, Munich), on the other. As in the study of Wagner & Jonkers (2011), we assume a delay of two years between funding and output and accordingly use OECD funding data for the period 2001-2011.[3] Because we have a time-series of observations, the publication year of the papers was added to the model as a third independent variable.

**Methods**

FWCI is a relative measure, whereas our independent variables are numbers of papers and US$ PPP. In order to avoid problems with this difference in the scale of the measurement, we use percentile classes of papers as dependent variables at 50%, 10%, and 1% of the most frequently cited papers, normalized with reference to the corresponding subject categories in WoS and publication years (see Table 1). Only papers with the document type "article" are considered. In the case of ties in citation numbers at the respective thresholds, the countries' papers are fractionally assigned to the percentile classes (Waltman and Schreiber, 2013). The resulting numbers were rounded off.

---

[1] We follow Wagner and Jonkers (2017) and use GBARD (OECD, 2017, p. 2; cf. Luwel, 2004, p. 327), and not Gross Expenditure in R&D (GERD) or Higher-Education Expenditure in R&D (HERD). GERD includes business funding and HERD excludes the Academies.
[2] These seven countries are: Argentina, China, Romania, the Russian Federation, Singapore, South Africa, and Taiwan.
[3] Funding data is retrieved from the OECD online at http://stats.oecd.org/Index.aspx?DataSetCode=MSTI_PUB .

**Table 1**. Key numbers for the variables included in the regression models

| Variables | Mean | Standard deviation | Minimum | Maximum |
|---|---|---|---|---|
| top 1% papers | 335.92 | 755.23 | 0 | 5457 |
| top 10% papers | 3340.31 | 7210.89 | 9 | 49855 |
| top 50% papers | 15873.4 | 31251.06 | 52 | 212857 |
| International collaboration | 11658.41 | 17642.31 | 85 | 131331 |
| Expenditure (US Dollars, Millions) | 8390.895 | 21975.7 | 24.66 | 164292 |
| Publication year | 2007.65 | 3.47 | 2002 | 2013 |

Three independent variables are used: 1) The annual number of internationally co-authored papers for each country; 2) government budget allocations for R&D (GBARD) in the publication year $y - 2$ assuming expenditures to show output with two year lag; 3) the publication year of the papers.

The dependent variables are count variables concerned by overdispersion, so we perform negative binomial regression models (Long & Freese, 2006). The regression models are based on n = 417 observations of "publication year x expenditure (country)" combinations. The countries are considered between 1 and 12 times in the analyses (on average 11 times). The cluster option in Stata is used to correct for this dependency in the data (Hilbe, 2014). We tested for multi-collinearity of the independent variables, but found—according to the guidelines of Acock (2016)—scarcely any hint of a multi-collinearity problem.

**Results**

The results of the models show that the coefficients for international collaboration and expenditure are close to zero (see Table 2).

**Table 2.** Coefficients and $t$ statistics from three negative binomial regression models

|  | (1) top 50% papers | (2) top 10% papers | (3) top 1% papers |
|---|---|---|---|
| International collaboration | 0.00*** | 0.00*** | 0.00*** |
|  | (4.98) | (5.13) | (5.30) |
| Expenditure | -0.00 | -0.00 | -0.00** |
| (US Dollars, Millions) | (-1.10) | (-1.95) | (-2.62) |
| Publication year | -0.01 | -0.01 | 0.03* |
|  | (-0.71) | (-0.43) | (2.25) |
| Constant | 21.97 | 15.29 | -51.49 |
|  | (1.12) | (0.72) | (-2.09) |
| Observations | 417 | 417 | 417 |

$t$ statistics in parentheses

* $p < 0.05$, ** $p < 0.01$, *** $p < 0.001$

In order to interpret the results of the regression models, Table 3 shows average marginal effects. These effects are changes in the dependent variable when the independent variable is increased by one unit (and the other independent variables are set to the mean value).

**Table 3**. Marginal effects with one unit change in the independent variable (+1)

|  | Change | Confidence interval | |
|---|---|---|---|
| *top 50% papers* | | | |
| International collaboration | 0.742 | 0.495 | 0.988 |
| Expenditure (US Dollars, Millions) | -0.177 | -0.471 | 0.117 |
| *top 10% papers* | | | |
| International collaboration | 0.161 | 0.104 | 0.218 |
| Expenditure (US Dollars, Millions) | -0.049 | -0.093 | -0.004 |
| *top 1% papers* | | | |
| International collaboration | 0.016 | 0.010 | 0.022 |
| Expenditure (US Dollars, Millions) | -0.005 | -0.009 | -0.002 |

The results can be interpreted as follows: on average, an increase of funding by one US$ million PPP decreases the expected number in the 50%, 10%, and 1% most-highly cited papers by 0.18, 0.05, and 0.01 papers, respectively. On average, the addition of one internationally co-authored paper increases the expected numbers of papers in these categories with 0.7, 0.2, and 0.02, respectively. Note that these latter values are much higher than the statistically expected ones in the three percentile classes (0.5, 0.1, and 0.01, respectively).

**Conclusions and discussion**

We confirm findings that international collaboration has a statistically significant and positive effect on the citation impact of nations. Increases in government funding, however, tend to have a negative or negligible effect on citation impact. Increased government funding seems not to be absorbed by authors and institutions who produce more highly cited papers, but by others like those at the bottom of the hierarchy, the bureaucracy, or it dissipates in the organization.

Our conclusions are "on average:" some nations appear to be more effective in turning funding into citation impact than others—several small nations punch above their weight in impact relative to spending. It may well be that the influence of government funding for some (e.g., capital-intensive) domains is different from others. Leydesdorff & Wagner (2008) found large differences in the price (in US$) per paper among nations. Some countries may have more slack and bureaucracy in the organization of the sciences than others (cf. Shelton & Leydesdorff, 2012). However, our results suggest support for the thesis that international collaboration in science has become a source of credit accumulation (Wagner, 2008).

Policy towards R&D investment has been based on consensus that one needs more science to thrive technology-based growth (e.g., Coccia, 2010; Grupp, 1995). The underlying assumption has been that national agents are able to appropriate the benefits of national public spending. This research suggests that the links between funding and outputs are disturbed by the rise of an international class of researchers who are decoupled from a national base. This new configuration has implications for accounting for the benefits of public funding, which requires additional inquiry and discussion.

**Acknowledgements**

The bibliometric data used in this paper are from an in-house database developed and maintained by the Max Planck Digital Library (MPDL, Munich) and derived from the Science Citation Index Expanded (SCI-E), Social Sciences Citation Index (SSCI), Arts and Humanities Citation Index (AHCI) prepared by Clarivate Analytics, formerly the IP & Science business of Thomson Reuters (Philadelphia, Pennsylvania, USA).

Wagner, C. S., Whetsell, T. A., & Leydesdorff, L. (2017). Growth of international collaboration in science: revisiting six specialties. *Scientometrics*, *110*(3), 1633-1652.

Wagner, C. S., & Jonkers, K. (2017). Open countries have strong science. *Nature News, 550*(7674), 32.

Waltman, L., & Schreiber, M. (2013). On the calculation of percentile-based bibliometric indicators. *Journal of the American Society for Information Science and Technology, 64*(2), 372-379.9